\documentclass[showpacs,preprintnumbers,amsmath,amssymb,floatfix]{revtex4}

\usepackage{graphicx}
\usepackage{epsfig}
\usepackage{amsfonts}
\usepackage{amssymb}
\usepackage{epsf}
\newcommand{\insertplot}[5]{\begin{figure}
 \hfill\hbox to 0.05in{\vbox to #5in{\vfill
 \inputplot{#1}{#4}{#5}}\hfill}
 \hfill\vspace{-.1in}
 \caption{#2}\label{#3}
 \end{figure}}
\newcommand{\inputplot}[3]{
 \special{ps: plotfile #1}
}

\newcounter{fig}

\begin{document}

\title{Quadrupole Moments of Rapidly
Rotating Compact Objects\\ in Dilatonic Einstein-Gauss-Bonnet Theory}

\author{
{\bf Burkhard Kleihaus, Jutta Kunz, and Sindy Mojica}
}
\affiliation{
{Institut f\"ur Physik, Universit\"at Oldenburg,
D-26111 Oldenburg, Germany}
}
\date{\today}
\begin{abstract}
We consider rapidly rotating black holes and neutron stars
in dilatonic Einstein-Gauss-Bonnet (EGBd) theory
and determine their quadrupole moments,
which receive a contribution from the dilaton.
The quadrupole moment of EGBd black holes 
can be considerably larger than the Kerr value.
For neutron stars, 
the universality property of the $\hat I$-$\hat Q$ relation
between the scaled moment of inertia and the scaled
quadrupole moment appears to extend
to EGBd theory.
\end{abstract}
\maketitle

\section{Introduction}

With numerous high precision telescopes and satellite missions 
operating already or becoming operational within the next decade
Einstein's theory of general relativity will be tested with
increasing accuracy, in particular, in the strong gravity regime
(see, e.g., the reviews~\cite{Psaltis:2008bb,Kramer:2009zza,Gair:2012nm,Yunes:2013dva,Will:2014xja}).

From a theoretical point of view general relativity 
is expected to be superseded by a new theory of gravity,
which on the one hand will not be suffering from those 
pathological singularities present in many
solutions of General Relativity (GR),
and, on the other hand, will be compatible with quantum theory.

String theory, as a leading candidate
for a unified theory of the fundamental interactions and
a quantum theory of gravity, for instance, predicts the
presence of higher curvature corrections to Einstein gravity
as well as the existence of further fields.
The low energy effective theory deriving from heterotic
string theory contains as basic ingredients a Gauss-Bonnet
term and a dilaton field \cite{Gross:1986mw,Metsaev:1987zx}.

The effects of the presence of such terms on the properties
of compact astrophysical objects can be strong,
as has been shown for black holes
\cite{Kanti:1995vq,Pani:2009wy,Kleihaus:2011tg,Pani:2011gy,Ayzenberg:2014aka}
and neutron stars \cite{Pani:2011xm},
which have been studied in a truncated version of this
theory, named dilatonic Einstein-Gauss-Bonnet (EGBd) theory.
Moreover, EGBd theory allows for wormhole solutions
without the necessity of introducing exotic fields
\cite{Kanti:2011jz,Kanti:2011yv}.

In this letter we focus on rapidly rotating compact objects
in EGBd theory and compare their properties with those of
their Einstein gravity counterparts.
In particular, we here determine the quadrupole moments of
rapidly rotating black holes and neutron stars.

Black holes in EGBd theory can slightly exceed the Kerr bound,
$J/M^2 =1$, given by the scale invariant ratio
of the angular momentum $J$ and the square of the mass $M$ 
in appropriate units
\cite{Kleihaus:2011tg}. 
The innermost-stable-circular-orbits (ISCOs) can
differ from the respective Kerr values
by a few percent in the case of
slow rotation \cite{Pani:2009wy}
and up to 10\% for rapid rotation \cite{Kleihaus:2011tg}.
Similarly, the orbital frequencies exhibit the largest deviations
from the Kerr frequencies for rapid rotation,
which can be as large as 60\%.
Such large effects might be observable 
for astrophysical black holes.

For neutron stars, on the other hand, recent
calculations showed, that for slowly rotating neutron stars
universal relations hold between the scaled moment
of inertia $\hat I$, the Love number, and the scaled quadrupole moment 
$\hat Q$ in Einstein gravity \cite{Yagi:2013bca,Yagi:2014bxa,Yagi:2014qua}.
Subsequently, the $\hat I$-$\hat Q$ relation 
was generalized to the case of rapidly rotating neutron stars,
where for fixed rotation parameter $j=J/M^2$
again only little dependence on the equation of state 
was seen \cite{Chakrabarti:2013tca},
after appropriately scaled quantities were considered
\cite{Doneva:2013rha}.

In the following we briefly recall EGBd theory
and obtain the quadrupole moment for compact objects
in this theory. We then
discuss the quadrupole moments of rapidly rotating black holes
and show that they may deviate from their Kerr counterparts by 20\% and more.
Subsequently, we present rapidly
rotating neutron stars in EGBd theory
and discuss their $\hat I$-$\hat Q$ relation.
We conclude, that just like in Einstein gravity
the properly scaled dimensionless quantities
exhibit little dependence on the equation of state.

\section{Einstein-Gauss-Bonnet-dilaton Theory}

As motivated by the low-energy heterotic string theory
\cite{Gross:1986mw,Metsaev:1987zx}
we consider the following low-energy effective action
\cite{Mignemi:1992nt,Kanti:1995vq,Chen:2009rv,Kleihaus:2011tg}
\begin{eqnarray}  
S=\frac{c^4}{16 \pi G}\int d^4x \sqrt{-g} \left[R - \frac{1}{2}
 (\partial_\mu \phi)^2
 + \alpha  e^{-\gamma\phi} R^2_{\rm GB}   \right],
\label{act}
\end{eqnarray} 
where $G$ is Newton's constant, $c$ is the speed of light,
$\phi$ denotes the dilaton field
with coupling constant $\gamma$, $\alpha $ is a numerical
coefficient given in terms of the Regge slope parameter,
and
$R^2_{\rm GB} = R_{\mu\nu\rho\sigma} R^{\mu\nu\rho\sigma}
- 4 R_{\mu\nu} R^{\mu\nu} + R^2$ 
is the Gauss-Bonnet (GB) term.
We have chosen the dilaton coupling constant according to its
string theory value, $\gamma=1$ .

To construct rotating compact objects
we employ the Lewis-Papapetrou line element \cite{Wald:1984rg} 
for a stationary, axially symmetric spacetime with
two Killing vector fields $\xi=\partial_t$,
$\eta=\partial_\varphi$. 
In terms of the spherical coordinates $r$ and $\theta$, 
the isotropic metric reads \cite{Kleihaus:2000kg}
\begin{equation}
ds^2 = g_{\mu\nu}dx^\mu dx^\nu
= -c^2 e^{2 \nu_0}dt^2
      +e^{2(\nu_1-\nu_0)}\left(e^{2 \nu_2}\left[dr^2+r^2 d\theta^2\right] 
       +r^2 \sin^2\theta
          \left(d\varphi-\omega c dt\right)^2\right) ,
\label{metric} 
\end{equation}
where $\nu_0$, $\nu_1$, $\nu_2$ and $\omega$ 
are functions of $r$ and $\theta$ only.

In the asymptotic region the metric and dilaton functions possess the expansion
\begin{eqnarray}
\nu_0 & = & -\frac{M}{r} + \frac{D_1 M}{3r^3} - \frac{M_2}{r^3} P_2(\cos\theta) +{\cal O}(r^{-4}) ,
\label{exnu0}\\
\nu_1 & = & \frac{D_1}{r^2} +{\cal O}(r^{-3}) ,
\label{exnu1}\\
\nu_2 & = & -\frac{4 M^2+16 D_1+ q^2}{8 r^2}\sin^2\theta +{\cal O}(r^{-3}) ,
\label{exnu2}\\
\omega & = & \frac{2 J}{r^3}  +{\cal O}(r^{-4}) ,
\label{exom}\\
\phi & = & \frac{q}{r}  +{\cal O}(r^{-2}) ,
\label{exdil}
\end{eqnarray}
where $P_2(\cos\theta)$ is a Legendre polynomial,
$M$ is the mass, $J$ is the angular momentum,
$q$ is the dilaton charge,
and $D_1$ and $M_2$ are further expansion constants.

To extract the quadrupole moment of the compact solutions,
we follow Geroch and Hansen \cite{Geroch:1970cd,Hansen:1974zz}
(see also \cite{Hoenselaers:1992bm,Sotiriou:2004ud})
and obtain
\begin{equation}
Q 
=  -M_2 +\frac{4}{3}\left[\frac{1}{4}+\frac{D_1}{M^2}
+\frac{q^2}{16M^2}\right] M^3 .
\label{Q}
\end{equation}
We note, that there is no explicit contribution from the GB term,
since this term decays sufficiently fast.
The dilaton field enters only via the Coulomb-like term ${q}/{r}$, 
which coincides with the analogous term in Einstein-Maxwell theory to
lowest order in the expansion. 
Thus the expressions for the quadrupole moment are completely analogous.
In the vacuum limit, when the dilaton charge vanishes,
the quadrupole moment $Q$ reduces
to the expression in \cite{Pappas:2012ns} (up to an overall sign),
with $b=D_1/M^2$.
On the other hand, in the static limit the solution is spherically symmetric.
In this case $M_2=0$ and $4M^2 + 16 D_1 + q^2 =0$, 
and the quadrupole moment vanishes.

\section{Rotating Black Holes}

\begin{figure}[h!]
\begin{center}
\mbox{
\includegraphics[height=.25\textheight, angle =0]{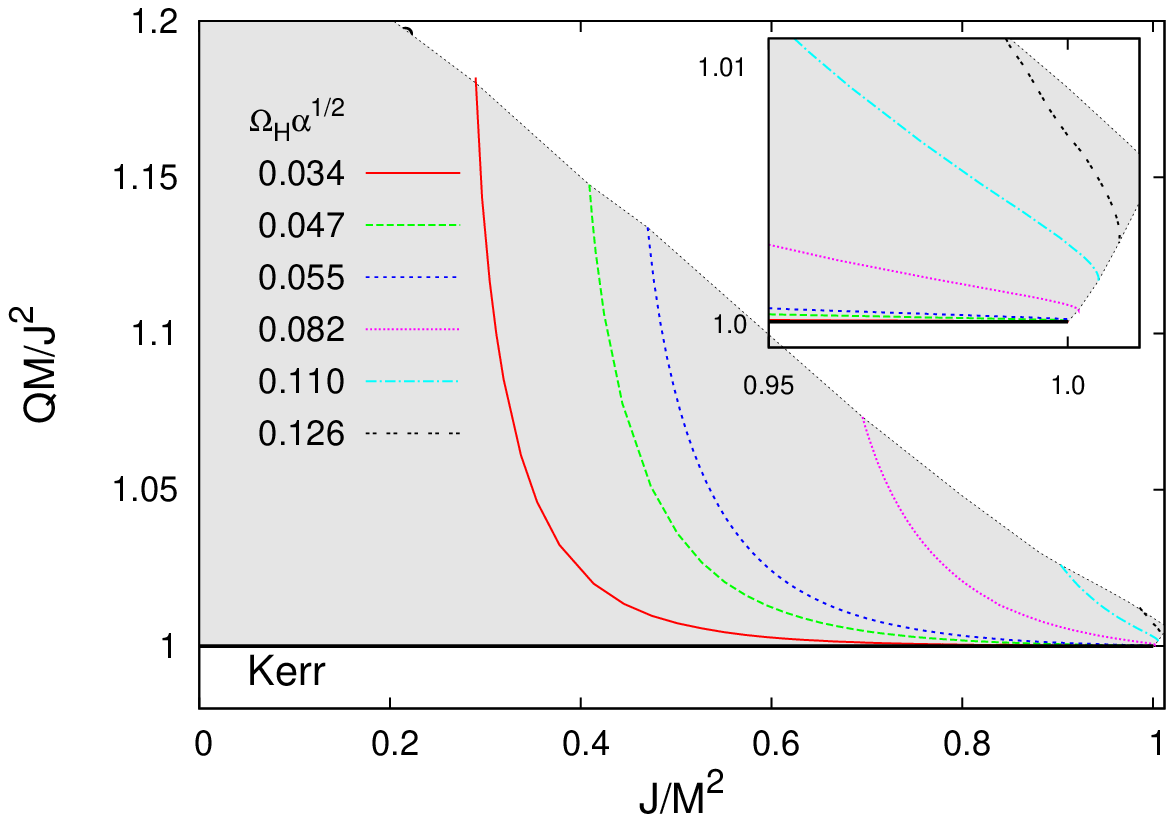}
\includegraphics[height=.25\textheight, angle =0]{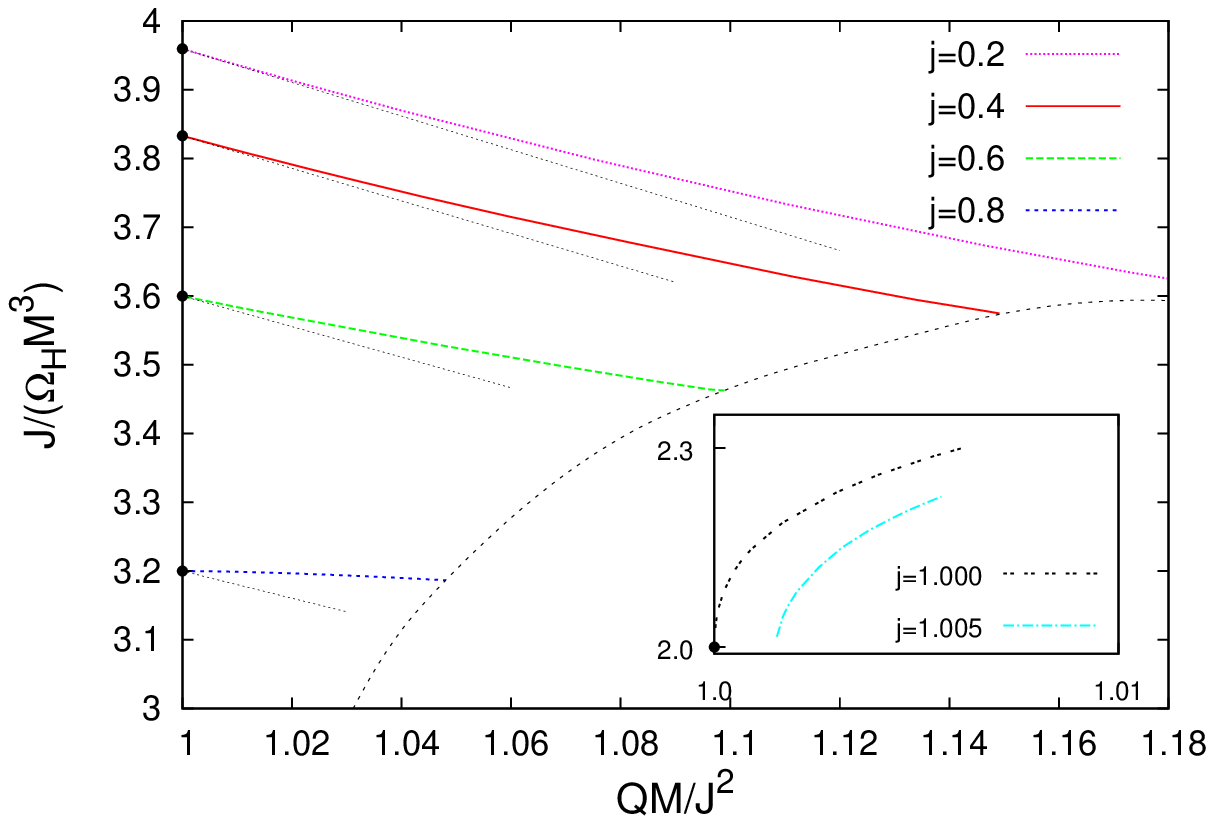}
}
\end{center}
\vspace{-0.5cm}
\caption{
(a) The scaled quadrupole moment $\hat{Q}=Q M/J^2$ 
is shown versus the scaled angular momentum $j=J/M^2$ 
for fixed values of the dimensionless
horizon angular velocity $\Omega_{\rm H}\alpha^{1/2}$.
The shaded area represents the domain of existence
of the EBGd black holes.
(b) The scaled moment of inertia
$\hat{I} = J/(\Omega_{\rm H} M^3)$ 
is shown versus the scaled quadrupole moment $\hat{Q}$
for fixed values of $j$. 
The dots indicate the corresponding values of the
Kerr black holes. 
The dotted curve corresponds to the 
boundary of the domain of existence given by the critical solutions.
The straight dotted lines represent the perturbative results
of \cite{Ayzenberg:2014aka}.
\label{fig1}
}
\end{figure}

The domain of existence of the EGBd black hole solutions \footnote{The
EGBd black holes are obtained with the ansatz 
\cite{Kleihaus:2011tg} for the line element
$
ds^2 = -e^{2\mu_0} dt^2 + e^{2\mu_1}\left(dr^2+r^2 d\theta^2\right)
       +e^{2\mu_2} r^2 \sin^2\theta \left(d\varphi -\omega dt\right)^2 ,
$
i.e., $\nu_0=\mu_0, \nu_1= \mu_2+\mu_0, \nu_2 = \mu_1-\mu_2$.}
resides
within the boundaries formed by (i) the Kerr black holes,
(ii) the critical black holes, where a radicant vanishes 
\cite{Kanti:1995vq,Kleihaus:2011tg},
and (iii) the set of extremal black holes with $j>1$,
which have temperature $T=0$, a regular metric, but a dilaton field
that diverges on the horizon at the poles
\cite{Kleihaus:2011tg}.
In Fig.~\ref{fig1}a the shaded area represents the domain of existence,
with (i) the lower boundary,
(ii) the upper boundary, and (iii) the right lower boundary in the inset.

In Fig.~\ref{fig1}a we present 
the scaled quadrupole moment $\hat{Q}=Q M/J^2$ of the 
rotating black hole solutions
versus the scaled angular momentum $j=J/M^2$
for several values of the dimensionless
horizon angular velocity $\Omega_{\rm H}\alpha^{1/2}$.
The curves are obtained, by fixing the value of
$\Omega_{\rm H}$ and the value of $\alpha$
and by varying the size of the black hole.

For fixed $\alpha$ the range of the scaled 
area $a=A/M^2$ for EGBd black holes
is largest in the static case and decreases to zero
for the maximal value of $j$ \cite{Kleihaus:2011tg}.
This is reflected in the range of the scaled quadrupole moment $\hat{Q}$,
which is likewise largest in the limit of slow rotation,
and decreases with increasing $j$.
Thus for slow rotation deviations from
the Kerr value of up to 20\% and more are seen.
Moreover, the solutions with $j>1$,
not present in General Relativity (GR), always have $\hat Q>1$.

In Fig.~\ref{fig1}b we exhibit the scaled moment of inertia
$\hat{I} = J/(\Omega_{\rm H} M^3)$ 
versus the scaled quadrupole moment $\hat{Q}$
for fixed values of $j$. 
For $j\le 1$ the families of solutions with fixed $j$ start
at the respective Kerr values, 
$\hat{I}_{\rm Kerr}=2(1+\sqrt{1-j^2})$ and $\hat{Q}_{\rm Kerr}=1$,
indicated by the dots on the axis.
They end at the boundary (ii) given by the critical solutions
and represented by the dotted curve.
The inset again shows the region $j>1$. 
The figure also contains the perturbative results derived
in \cite{Ayzenberg:2014aka} for small $\alpha$ and small $j$.

Clearly, the $\hat I$-$\hat Q$ relation for black holes depends
on the value of the scaled angular momentum $j$.
However, for black holes there is no dependence of this
relation on the coupling parameter $\alpha$.
This is a consequence of the invariance of $\hat{I}$,
$\hat{Q}$, $j$ and the EGBd equations under the scaling transformation
$r \to \lambda r$, $\omega \to \omega/\lambda$, 
and $\alpha \to \lambda^2 \alpha$.

The absolute value of the
scaled dilaton charge $q/M$ decreases monotonically
with decreasing scaled quadrupole moment $\hat Q$.
In the limit $\hat{Q} \to 1$ the scaled
dilaton charge $q/M$ vanishes.
Therefore the scaled moment of inertia $\hat{I}$ 
approaches $\hat{I}_{\rm Kerr}$ in this limit.
Its dependence on the value of the scaled angular momentum $j$
is only weak.

\section{Rotating Neutron Stars}

Rapidly rotating neutron stars have been studied
extensively in GR (see, e.g., \cite{Stergioulas:2003yp,Friedman:2013}).
Here we address rapidly rotating neutron stars in EGBd theory
\cite{Kleihaus:2011tg}.
Considering rigid rotation,
the four-velocity has the form
$U^\mu = \left( u, 0, 0, \Omega u\right) $
with angular velocity $\Omega$.
The normalization condition $U^\mu U_\mu = -1$ then yields
\begin{equation}
u^2 = \frac{e^{-2\nu_0}}{1-(\Omega-\omega)^2 r^2\sin^2\theta e^{2\nu_1-4\nu_0}}  .
\label{u2}
\end{equation}
The differential equations for the pressure $P$ 
and the energy density $\epsilon$ are obtained
from the constraints $\nabla_\mu T^{\mu\nu}= 0$,
\begin{equation}
\frac{\partial_r P}{\epsilon +P}  =  \frac{\partial_r u}{u} 
 , \ \ \
\frac{\partial_\theta P}{\epsilon +P}  =  \frac{\partial_\theta u}{u} .
\label{dpres}
\end{equation}

These equations have to be supplemented by an equation of state (EOS), 
$\epsilon = \epsilon(P)$ (or $P = P(\epsilon)$).
For a polytropic EOS,
$
P = P_0 \Theta^{N+1}, \ \epsilon = N P + \rho_0 \Theta^N
$, with central pressure $P_0$ and central density $\rho_0$.
Substitution in Eq.~(\ref{dpres}) yields
$
\Theta = c_0 u - {\rho_0}/{P_0(N+1)} 
$,
where $c_0$ is an integration constant,
which we express as
$
c_0={\rho_0}/{\sigma P_0(N+1)}
$
to obtain the more convenient expression
\begin{equation}
\Theta =\frac{\rho_0}{\sigma P_0(N+1)}(u -\sigma) \ .
\label{theta1}
\end{equation}
The boundary of the star is defined by $P=P_b=0$, 
implying $\Theta_b=0$ and consequently $u_b =\sigma$. 
Outside the star $\Theta=0$.

\begin{figure}[t!]
\begin{center}
\mbox{
\includegraphics[height=.25\textheight, angle =0]{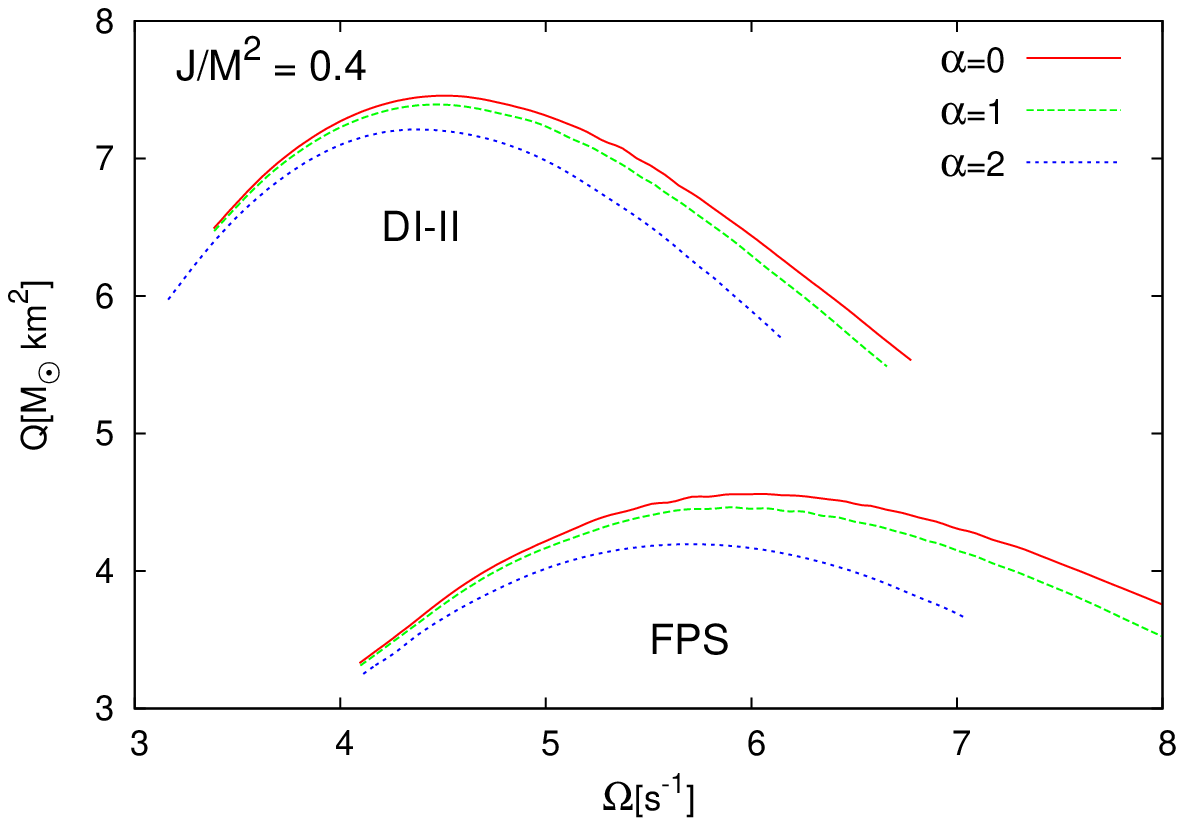}
\includegraphics[height=.25\textheight, angle =0]{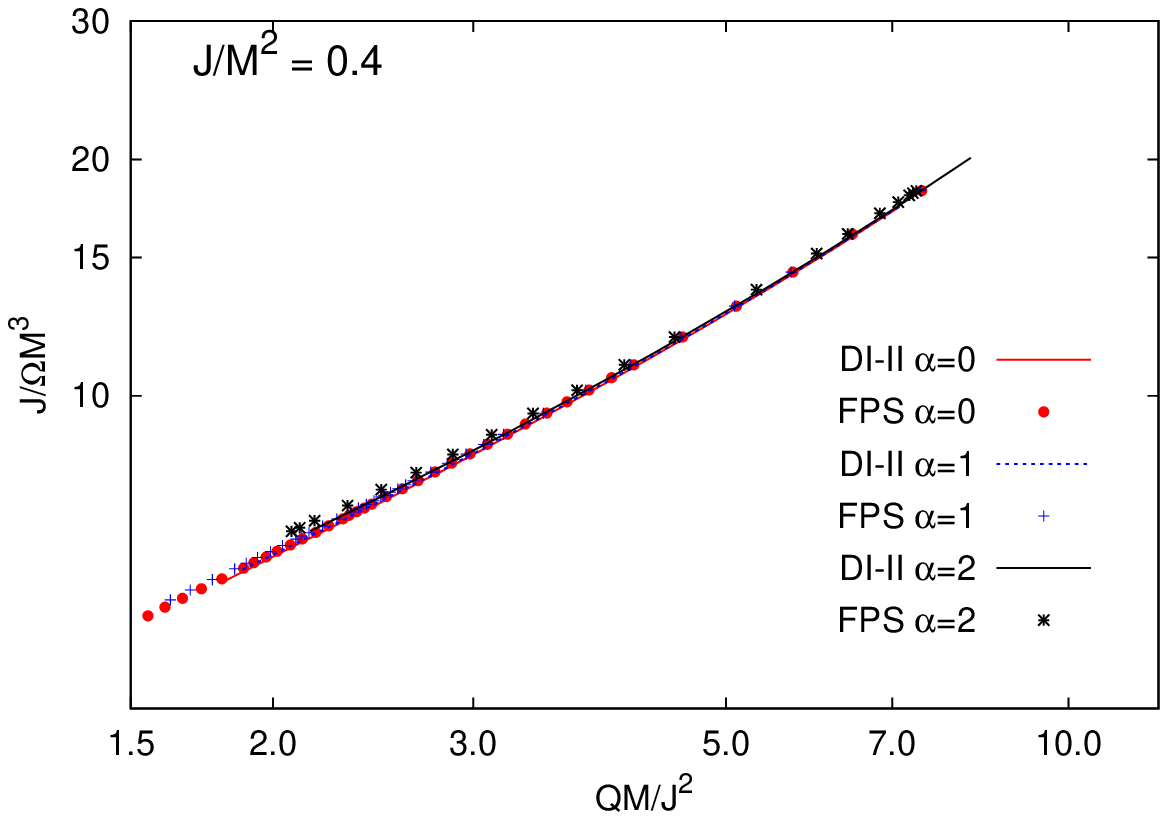}
}
\end{center}
\vspace{-0.5cm}
\caption{
(a) The quadrupole moment $Q$ in units of $M_\odot \cdot {\rm km}^2$
versus the angular velocity $\Omega$ in Hz
for fixed scaled angular momentum $j=0.4$,
GB couplings $\alpha=0$, 1 and 2,
the polytropic EOS DI-II \cite{Diaz-Alonso:1985}
and the approximated FPS EOS \cite{Haensel:2004nu}.
(b) The scaled moment of inertia $\hat{I}$ versus
the scaled quadrupole moment $\hat{Q}$
for the same set of solutions.
}
\label{fig2} 
\end{figure}

In Fig.~\ref{fig2}a
we exhibit the quadrupole moment $Q$ 
in units of $M_\odot \cdot {\rm km}^2$
versus the angular velocity $\Omega$ in Hz
for a family of neutron stars 
for a fixed scaled angular momentum of $j=0.4$,
the GB couplings $\alpha=0$, 1 and 2,
and two well-known equations of state.

The first EOS corresponds to a polytropic EOS with $N=0.7463$,
taken from \cite{Diaz-Alonso:1985} and denoted by DI-II,
which is widely used in neutron star physics
and also employed for neutron stars in scalar-tensor theory
\cite{Damour:1993hw,Doneva:2013qva},
while the second EOS corresponds
to an approximation\footnote{The analytical fit \cite{Haensel:2004nu}
to the FPS EOS is approximated by a fit to a polytropic
EOS.} to the FPS EOS
\cite{Haensel:2004nu}.
Clearly, the two equations of state give
rise to rather different quadrupole moments,
whereas the dependence on the GB
coupling $\alpha$ is only moderate,
while increasing with increasing $\alpha$.

To address the $\hat I$-$\hat Q$ relation for neutron stars
we exhibit in Fig.~\ref{fig2}b 
$\hat{I}$ versus $\hat{Q}$
for a fixed value of the scaled angular momentum, $j=0.4$.
Again, we employ
the polytropic EOS DI-II \cite{Diaz-Alonso:1985}
and the analytical fit to the FPS EOS
\cite{Haensel:2004nu}.
 
Comparison shows, that for a given value of $\alpha$
and a given value of $j$
the dependence on the EOS is only weak,
but increasing with increasing $\alpha$.
Moreover, the $\alpha$-dependence is the stronger
the smaller $j$.
We did not go beyond $\alpha=2$, since the 
strongest current observational upper bound
on $\alpha$, obtained from low-mass X-ray binaries,
is slightly less than this value (in the units employed)
\cite{Yagi:2012gp,Ayzenberg:2014aka}.

\section{Conclusion}

Following the procedure of Geroch and Hansen
\cite{Geroch:1970cd,Hansen:1974zz}
we have derived the spin induced quadrupole moment
for rapidly rotating compact objects in EGBd theory.
The scaled quadrupole moment of black holes can deviate
by 20\% and more from the corresponding Kerr value
for a given value of the scaled angular momentum $j$.
Their $\hat I$-$\hat Q$ relation has no deplicit dependence
on the Gauss-Bonnet coupling $\alpha$.

For neutron stars the presence of the matter breaks the
scale invariance of the $\hat I$-$\hat Q$ relation.
We now observe a dependence on $\alpha$, which is however
weak in its allowed range.
While the unscaled quantities depend quite significantly
on the equation of state, the scaled quantities satisfy
to good approximation universal relations,
which depend only on the scaled angular momentum $j$, 
in EGBd theory like in GR.

We are currently extending the calculations to quark stars,
to see whether the universal relations hold also
in this case \cite{Yagi:2013bca} for EGBd theory.
Unlike the case of Chern-Simons theory
also studied in \cite{Yagi:2013bca}, however,
the $\hat I$-$\hat Q$ relations do not differ
significantly for EGBd theory and GR.
Here other observables including ISCOs and orbital periods
seem more promising handles to test the viability of EGBd theory.
 
Finally, another type of compact object can be obtained in
EGBd theory, namely wormholes, which may possibly correspond to
black hole foils \cite{Damour:2007ap}.
We expect that the static EGBd wormholes \cite{Kanti:2011jz,Kanti:2011yv}
can be generalized to rapidly rotating wormholes, while retaining
their linear stability with respect to changes of the size of their throat.
It will be interesting to extract their quadrupole moments
and compare them with the quadrupole moments of the EGBd black holes.

\newpage
\vspace{0.2cm}
{\sl Acknowledgement}

{We gratefully acknowledge discussions with
Tibault Damour, Norman G\"urlebeck and Eugen Radu
as well as support by the DFG
within the Research Training Group 1620 ``Models of Gravity''.}

\end{document}